\documentclass[aps,pra,showpacs,notitlepage]{revtex4-1}

\usepackage{graphicx}

\begin{document}

\title{Why the ``classical'' explanation of weak values by Ferrie and Combes does not work:\\
a comment on Phys. Rev. Lett. 113, 120404 (2014)}

\author{Holger F. Hofmann}
\email{hofmann@hiroshima-u.ac.jp}
\affiliation{
Graduate School of Advanced Sciences of Matter, Hiroshima University,
Kagamiyama 1-3-1, Higashi Hiroshima 739-8530, Japan}

\author{Masataka Iinuma}
\email{iinuma@hiroshima-u.ac.jp}
\affiliation{
Graduate School of Advanced Sciences of Matter, Hiroshima University,
Kagamiyama 1-3-1, Higashi Hiroshima 739-8530, Japan}

\author{Yutaka Shikano}
\email{yshikano@ims.ac.jp}
\affiliation{ Institute for Molecular Science, 38 Nishigo-Naka, Myodaiji, Okazaki 444-8585, Japan}
\affiliation{Institute for Quantum Studies, Chapman University,  1 University Dr., Orange, CA 92866, USA}

\begin{abstract}
In Phys. Rev. Lett. 113, 120404 (2014), Ferrie and Combes claim that weak values could be a statistical artifact caused by correlations between the disturbance and the post-selection condition imposed on the output. In this comment, we show that the disturbance caused by a weak measurement is sufficiently low to definitely rule out the model proposed by Ferrie and Combes. 
\end{abstract}

\maketitle

In their recent Letter, Ferrie and Combes claim that weak values could be a statistical artifact caused by correlations between the disturbance and the post-selection condition imposed on the output \cite{Fer14}. The basis of this claim is a classical statistical model unrelated to weak measurements, where the disturbance is maximally correlated with the measurement errors. Unfortunately, the authors fail to examine the magnitude of the disturbance in weak measurements. In fact, it can be shown that the disturbance in properly performed weak measurements is always low enough to completely rule out the effect described by Ferrie and Combes. We therefore conclude that the criticism of Ferrie and Combes is unfounded, and that the claim that weak values can be explained as a ``purely statistical feature of pre- and postselection with disturbance'' cannot be upheld. 

The disturbance caused by a weak measurement can be quantified easily by using the measurement operators $M_q$ given in Eq.(7) of the paper. Ironically, the authors have omitted the term that describes the back-action in their Eqs.(9) and (10), 
probably because it is proportional to the square of the measurement strength $\lambda$ and is therefore negligibly small in the weak measurement  limit (see \cite{Hof10} for a more detailed discussion). As a consequence, the sum over the results with $s=+1$ and the results with $s=-1$ in Eq.(10) describe a completely undisturbed output state of $\rho$. Thus the weak measurement theory presented by the authors themselves is disturbance free and should have a bit-flip probabiliy of $p=0$ in Eq.(22). This seems to contradict the  claim that ``One can even choose the $p$ here so that it is identical to effective $p$ from the fully quantum calculation.'' since the ``fully quantum calculation''  presented by the authors results in $p=0$. 

One might object that the omission of the disturbance proportinal to $\lambda^2$ is just an approximation, and that its justification should be re-examined based on the insight that post-selection might enhance the effects of the disturbance. However, a closer analysis of the present scenario answers this possible objection. From Eqs. (21, 23), it follows that the average value of $p$ is equal to $1-\delta$, so the minimal disturbance of a weak measurement is given by  $(1-\delta)=\lambda^2$.  The model of the authors requires a much larger disturbance of $(1-\delta)>\lambda$, so it can be ruled out as a possible explanation of weak measurements and weak value enhancements.

The letter of Ferrie and Combes certainly does concern an important practical problem: experimentalists must make sure that the disturbance caused by a weak measurement is sufficiently low to be neglected, even in the presence of post-selection. This requires that the post-selection probability does not show any significant change as a result of the weak measurement. Specifically, $P(\Phi)$ should satisfy $|\langle\Phi \mid \Psi \rangle|^2\gg(1-\delta)$. In the present scenario, $|\langle \Phi \mid \Psi \rangle|^2=0$, so the disturbance $(1-\delta)$ is the {\it only} contribution to the post-selected outcomes. This is clearly the opposite situation to a properly performed weak measurement. 

We would also like to point out that the effects of disturbance on weak measurement enhancements have been studied in quite some detail, e.g. in \cite{Iin11}. The results show that the disturbance caused by the measurement interaction tends to reduce the enhancement of the weak value, which clearly contradicts the assumption that the enhancement is caused by the disturbance. 

\newpage

In conclusion, Ferrie and Combes are mistaken about the possible explanation of weak measurement statistics - although it is possible to produce statistical artifacts that look like weak values, this effect is only possible if the disturbance by the measurement would be much larger than it actually is in properly performed weak measurements.


\begin{thebibliography}{xyz00}

\bibitem{Fer14}
C. Ferrie and J. Combes, Phys. Rev. Lett. {\bf 113}, 120404 (2014).

\bibitem{Hof10}
H. F. Hofmann, Phys. Rev. A {\bf 81}, 012103 (2010).

\bibitem{Iin11}
M. Iinuma, Y. Suzuki, G. Taguchi, Y. Kadoya, and H. F. Hofmann,
New J. Phys. {\bf 13}, 033041 (2011). 

\end{thebibliography}
\end{document}